\documentstyle[12pt,epsfig]{article}
\newlength{\dinwidth}
\newlength{\dinmargin}
\begin{document}
\newcommand{\be}{\begin{equation}}
\newcommand{\ee}{\end{equation}}
\newcommand{\ba}{\begin{eqnarray}}
\newcommand{\ea}{\end{eqnarray}}
\newcommand{\dy}{\Delta Y}
\newcommand{\tdm}[1]{\mbox{\boldmath $#1$}}
\newcommand{\idd}[1]{d^2\mbox{$\boldmath #1\,$}}
\newcommand{\as}{\alpha_s}
\def\la{\mathrel{\mathpalette\fun <}}
\def\ga{\mathrel{\mathpalette\fun >}}
\def\fun#1#2{\lower3.6pt\vbox{\baselineskip0pt\lineskip.9pt
\ialign{$\mathsurround=0pt#1\hfil##\hfil$\crcr#2\crcr\sim\crcr}}}
\titlepage
\begin{flushright}
IPPP/01/38, DCPT/01/76, \\
TSL/ISV-2001-0252      \\
October 19, 2001       \\
\end{flushright}
\begin{center}
\vspace*{2cm}
{\Large \bf
The non-forward BFKL amplitude and rapidity gap physics
}

\vspace*{1cm}
L.\ Motyka$^{a,b}$, A.D.\ Martin$^{c}$ and M.G.\ Ryskin$^{c,d}$ \\
\vspace*{0.5cm}
$^{a}$ High Energy Physics, Uppsala University, Sweden\\
$^{b}$ Institute of Physics, Jagellonian University, Krak\'{o}w, Poland \\
$^{c}$ Department of Physics and Institute for Particle Physics
Phenomenology,
University of Durham, Durham, DH1 3LE \\
$^{d}$ Petersburg Nuclear Physics Institute, Gatchina, St. Petersburg 188300, Russia\\
\end{center}

\vspace*{2cm}

\begin{abstract}

We discuss the BFKL approach to processes with large momentum
transferred through a rapidity gap. The Mueller and Tang scheme to
the BFKL non-forward parton-parton elastic scattering amplitude at
large $t$, is extended to include higher conformal spins. The
new contributions are found to decrease with increasing energy, as
follows from the gluon reggeisation phenomenon, and to vanish for
asymptotically high energies. However, at moderate energies and
high $|t|$, the higher conformal spins dominate the amplitude. We
illustrate the effects by studying the production of two high
$E_T$ jets separated by a rapidity gap at HERA energies. 
In a simplified framework, we find excellent 
agreement with the HERA photoproduction data once we incorporate the
rapidity gap survival probability against soft rescattering effects.
We emphasize that measurements of the analogous process in electroproduction
may probe different summations over conformal spins.
\end{abstract}

\newpage

\section{Introduction}

High energy hadronic scattering with large momentum transfer $|t|
\gg \Lambda^2 _{QCD}$ and rapidity $y \gg 1$ is an excellent
testing ground for perturbative QCD. The most interesting is the
case when colour is not exchanged in the interaction. Our
understanding of such processes (hard colour singlet exchange) is
based on the BFKL equation \cite{BFKL,LIPATOV,LIPAT1} which resums
the gluonic ladder diagrams in the leading logarithmic
approximation\footnote{Recently, also the next-to-leading
corrections have become available \cite{BFKLNLL}.}. The best known
processes in which the behaviour of the scattering amplitude may
be tested are elastic vector meson production
\cite{FR,BFLW,PSIPSI}, diffractive, proton dissociative $\gamma p$
scattering \cite{IVANOVGP} and events with gaps between jets in
the appropriate kinematic regime \cite{MT,CF,CFL,EMI,EIM}. Two main
approaches to determine non-forward BFKL amplitudes have been
proposed.

The first approach relies on the conformal symmetry of the leading logarithmic
BFKL equation, which permits an analytical solution of the problem
\cite{BFKLSOL,LIPAT1}.
The applications follow the Mueller and Tang \cite{MT} subtraction scheme
to obtain the elastic parton-parton scattering amplitude.
It is valid for an asymptotically large rapidity gap.
However, it is doubtful whether this asymptotic formula may be used for
the currently
available measurements. The main problem is, that the Mueller-Tang
cross-section for gaps between jets is, even for $y \sim 5$, much smaller
than the lowest order two-gluon exchange cross-section. Of course, one has
to include the gluon reggeization factor which suppresses the infrared sensitive
part of the two-gluon amplitude \cite{FR,BLLRW},
but still this contribution appears to be very important \cite{EMI,EIM}
for the region of $y$ probed in the current experiments \cite{TEVEXP}.

The other approach is based on numerical studies of the
non-forward BFKL equation \cite{PSIPSI,EMI,EIM}. The main advantage of
this method is that it does not require any restrictions imposed
on $y$ and takes into account all the available details of the
impact factors. An important ingredient of this framework is that
it is possible to go beyond the leading logarithmic approximation
by including some phenomenological modifications of the BFKL
kernel which are expected to resum a major part of the higher
order corrections, like the running of the coupling constant along
the ladder and the imposition of the consistency constraint
\cite{CC1,CC2,PSIPSI}.

The main purpose of this paper is to generalize the analytical
approach so that the BFKL parton scattering elastic amplitude can
be used at lower values of $y$. We will demonstrate that in this
case we have to include the higher conformal spin contributions
and so we are able to trace the phenomenon of gluon reggeization
in a representation given by the conformal eigenfunctions of the
BFKL kernel. 
Possible phenomenological effects of higher conformal spins 
in the forward BFKL amplitude have been discussed, for example, in 
\cite{RP1,RP2}.

From the experimental point of view, the process of interest (with
a large rapidity gap between two high $E_T$ jets) has been
observed at the Tevatron \cite{TEVEXP} and now data are becoming
available for the analogous process at HERA \cite{ZEUS,H1}. For
example, in Fig.~1 we show the production mechanism for the
diffractive process at HERA. Originally, it was claimed
\cite{TEVEXP} that the Tevatron observations strongly disagree
with the BFKL approach. However a closer study  shows that the
contradiction disappears when we allow for the effects of (i)
hadronisation, (ii) the survival probability of the rapidity gaps and
(iii) asymmetric non-asymptotic BFKL contributions. Numerically,
at these energies it was found at the partonic level \cite{EMI,EIM}
that the elastic parton-parton amplitude may be well described by
two reggeized gluon exchange. Indeed, as we will show in 
Section~2, the non-asymptotic (i.e. higher conformal spin) components may
be summed to give reggeized two gluon exchange. We note that the
conformal spin components contribute up to the value of the
conformal spin  $n \sim \sqrt{E_T/k_0}$, where the infrared
physical cutoff $k_0$ is driven by the size of the incoming state.

\begin{figure}[t]
\begin{center}
\mbox{\epsfig{figure=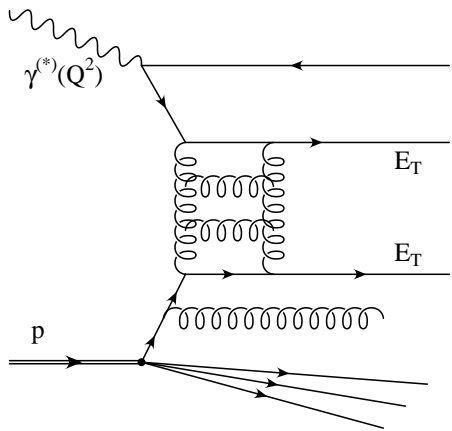,width=6.0cm}}
\end{center}
\caption{\small\em The Feynman diagram illustrating the mechanism
of photo- and electroproduction of two high $E_T$ jets separated by a rapidity gap.}
\end{figure}

In Section 3 we consider the corresponding process at HERA: that is
two high $E_T$ jets separated by a rapidity gap as shown in
Fig.~1. Here we have a second variable, the photon virtuality $Q^2$, which allows us to
change the size of the incoming $q\bar q$ state. In this way one
can study the effect of varying the number of conformal spin
components. We have attempted to summarize the essential points of
the analysis in a self-contained, and more physical, way when
describing the application in Section 3.

\section{Elastic parton-parton scattering amplitude}

The quark-quark elastic cross-section at high energies reads \be
{d\sigma_{qq} \over dt}(y) = {4\as^4 \over 81 \pi} |A(y,t)|^2,
\label{eq:dsdt} \ee where the amplitude $A(y,t)$ is \be A(y,t) =
\int  \idd{k} \idd{k}' f^q(\tdm{k}, \tdm{k}';y). \label{eq:ampl}
\ee

\subsection{Two gluon exchange}

In the $y \to 0$ limit, the BFKL amplitude reduces to its lowest
level approximation of the exchange of two elementary gluons, i.e.
\be f^q(\tdm{k}, \tdm{k}';0) = { \delta(\tdm k - \tdm k') \over
\tdm k^2 (\tdm{q}-\tdm{k})^2}. \label{eq:fkk} \ee The integral
over the tranverse momenta in eq.~(\ref{eq:ampl}) is infrared
divergent in this limit. One may, however, introduce an infrared
cutoff $k_0^2$ and modify the gluon propagators $1/ k^2 \to 1/(k^2
+k_0^2)$. Such cutoff has a physical origin. It may be related
either to hadron sizes or to the gluon propagation length in the
QCD vacuum. With such a substitution, (\ref{eq:ampl}) becomes \be
A(y=0,t) \simeq {2\pi \over q^2} \log (q^2 /k^2 _0) \label{eq:log}
\ee where $q$ is the momentum transfer. For the production of a
pair of high $E_T$ jets, we have $q\simeq E_T$. At not large
rapidities and large $|t|$, the most important effect of the BFKL
evolution is that the exchanged gluons become reggeized. This
accounts for a no-emission amplitude from a system of two gluons
in the colour singlet state, where one of the gluons is much
softer than the other. To a good approximation, it leads to an
additional factor multiplying the gluon propagator close to the
singular point \be {1 \over k^2 + {k_0 ^2}} \to {1 \over k^2+{k_0
^2}} \left( {k^2 \over q^2} \right)^z \label{eq:regg} \ee with $z
= 3\as y /(2\pi)$. Then for moderate $y$
\be
A(y,t)
\simeq
{2\pi \over q^2} \int_0 ^{q^2}  dk^2  {1 \over k^2+{k_0 ^2}} \left( {k^2 \over q^2} \right)^z.
\label{eq:regg2}
\ee
For $y \neq 0$  the integration in (\ref{eq:regg2})  may be safely
performed, and the  limit ${k_0 ^2} \to 0$ taken, to obtain \be A
(y,t) \; = \; { 2\pi \over q^2} {1\over z} \left[ 1 - \left( {k_0
^2} \over q^2 \right)^z \right] \to
 { 2\pi \over z q^2}.
\label{eq:univ} \ee We shall trace how the above expression arises
from the summation over the conformal spins.

\subsection{BFKL conformal components}

The LO BFKL expression for $f^q(\tdm{k}, \tdm{k}';y)$ is
\cite{BFKLSOL,LIPAT1}
\[
f^q(\tdm{k}, \tdm{k'};y) = {1\over (2\pi)^6}
\sum_{n=-\infty} ^\infty
\int d\nu \left\{ \,
{\nu^2 + n^2/4 \over [\nu^2 + (n-1)^2/4][\nu^2+(n+1)^2/4]} \right. \times
\]
\be 
\left. \qquad \qquad \exp [\omega_n(\nu) y]\; {I^1 _{n,\nu}}
(\tdm k,\tdm q)\, {I^2 _{n,\nu}}^* (\tdm k',\tdm q) \right\}
\label{eq:fbfkl} 
\ee 
where 
\be 
\omega_n(\nu) = {3 \as \over \pi} [
2\psi(1) - \psi(1/2+|n|/2+i\nu) - \psi(1/2+|n|/2-i\nu) ]
\label{eq:omega} 
\ee 
are the eigenvalues of the BFKL kernel. The
functions $I^s _{n,\nu}$ are constructed from the impact factors
$\Phi^s (\tdm k,\tdm q)$ and the eigenfunctions of the BFKL
kernel. To be precise \be I^s _{n,\nu} (\tdm k,\tdm q) = \Phi^s
(\tdm k,\tdm q) \int d^2\rho_1\, d^2\rho_2\,
E_{n,\nu}(\rho_1,\rho_2) \exp(ik\rho_1 + i(q-k)\rho_2)
\label{eq:is} \ee where the eigenfunctions take the form \be
E_{n,\nu}(\rho_1,\rho_2)  = {\left( \rho_1 - \rho_2 \over \rho_1
\rho_2 \right)}^h {\left(\left( \rho_1 - \rho_2 \over \rho_1
\rho_2 \right)^*\right)}^{\tilde h}. \label{eq:E} \ee Here we have
used the complex representation of transverse vectors $\tdm k$ and
$\tdm \rho$ (which reveals the conformal symmetry), that is $k =
k_x + i k_y$ and $\rho = \rho_x + i\rho_y$. The powers
$h=1/2+n/2+i\nu$ and $\tilde h = 1/2-n/2+i\nu$ are  the conformal
weights.

For parton scattering, the impact factors $\Phi^s (\tdm k,\tdm
q)$ are constant functions of  $\tdm k$. In other words for a
point-like parton we have $\rho_1=\rho_2$, and hence the
eigenfunctions (\ref{eq:E}), the impact factors (\ref{eq:is}) and
the amplitude $f^q$ of (\ref{eq:fbfkl}), vanish identically.
However one can not use the expansion  (\ref{eq:fbfkl}) over the
conformal eigenfunctions $E_{n,\nu}$ for coloured initial objects.
In such a case we face an infrared divergency in the integrations
over the $\rho_i$ which correspond to contributions proportional
to $\delta(\tdm k)$ or $\delta(\tdm q -\tdm k)$.  For a colourless
object these divergent terms are absent, since they are multiplied
by zero -- the total colour charge. Thus one has to consider the
parton spectators which compensate the colour charge of our active
quark. Assuming that these spectators are located at rather large
distances we may follow the Mueller-Tang
prescription, as was shown in \cite{BLLRW}.\\

The generalisation of Mueller-Tang prescription for arbitrary
conformal spins reads:\\[1ex]
\[
{\left( \rho_1 - \rho_2 \over \rho_1 \rho_2 \right)}^{h}
{\left(\left(\rho_1 - \rho_2 \over \rho_1 \rho_2 \right)^*\right)}^{\tilde h}
 \to {\left( \rho_1 - \rho_2 \over \rho_1 \rho_2 \right)}^{h}
{\left(\left( \rho_1 - \rho_2
\over \rho_1 \rho_2 \right)^*\right)}^{\tilde h} - \left({1 \over \rho_2}
\right)^{h}\left({1 \over \rho_2 ^*} \right)^{\tilde h}- \left({-1 \over
\rho_1} \right)^{h}\left({-1 \over \rho_1 ^*} \right)^{\tilde h} = \] \be
{\left( \rho_1 - \rho_2 \over \rho_1 \rho_2 \right)}^{h} {\left(\left(
\rho_1 - \rho_2 \over \rho_1 \rho_2 \right)^*\right)}^{\tilde h} - \left({1
\over |\rho_2|} \right)^{1+2i\nu} \left({\rho_2 \over |\rho_2|} \right)^{-n} -
 \left({1 \over |\rho_1|} \right)^{1+2i\nu} \left({-\rho_1 \over |\rho_1|}
 \right)^{-n} . \\[1em]
\label{eq:subtr}
\ee
Note the minus sign in the last term,
which will result in the cancellation of contributions with odd~$n$.  We set
$\Phi^s(\tdm q,\tdm k)=1$ for $s=1,2$ and substitute
(\ref{eq:subtr}) into (\ref{eq:is}) to obtain
\be
I^s _{n,\nu} (\tdm k,\tdm q) = -(2\pi)^3 \, i^n \,
[\delta(\tdm q - \tdm k) + (-1)^n \delta(\tdm k) ] \;
{1\over q} \left( {q \over 2 } \right)^{2i\nu}
{\Gamma(1/2+n/2 - i\nu) \over \Gamma(1/2 + n/2 + i\nu)}.
\label{eq:ifinal}
\ee
We insert the last result into eq.~(\ref{eq:fbfkl}) and then integrate over
$\tdm k$ and $\tdm k'$ to obtain the non-forward amplitude:
\be
A(y,t) =
{4 \over q^2} \sum_{m=-\infty} ^\infty
\int d\nu \left\{ \, {\nu^2 + m^2 \over [\nu^2 +
 (m-1/2)^2][\nu^2+(m+1/2)^2]} \; \exp [\omega_{2m}(\nu) y] \right\}.
\label{eq:final} 
\ee 
The last formula represents the desired
generalisation of the Mueller-Tang result for the quark-quark
elastic scattering amplitude. Note, that only contributions from
even conformal spins $n=2m$ are left in the sum.

\subsection{Sum over conformal spins}

The resulting amplitude has the following properties. For very
large $y$ all the components with $|m|>0$ get suppressed because
in this case $\omega_{2m}(\nu) < 0$. Then, indeed, it is enough to
retain only the leading term with $m=0$ (i.e. $n=0$), which gives
rise to an increasing part of the amplitude with the famous LO
BFKL intercept \be A(y,t) \sim y^{-3/2} \exp\left({12 \log{2} \;
\as \over \pi}\, y\right). \label{eq:lead} \ee However, for $y \to
0$ the expression is divergent, as expected from an inspection of
the two gluon exchange amplitude, due to high $1/|m|$ asymptotics
in the terms under the sum (\ref{eq:fbfkl}): \be \int d\nu \,
{\nu^2 + m^2 \over [\nu^2 + (m-1/2)^2][\nu^2+(m+1/2)^2]} = \left\{
\begin{array}{ll}
\displaystyle{\pi}                                           & {\rm for}\quad m=0, \\ \\
\displaystyle{{\pi\over |m|}\;{m^2-1/8 \over m^2-1/4}} \quad & {\rm for}\quad m \neq 0. \\
\end{array} \right.
\label{eq:y0} \ee For $y>0$ the divergence disappears due to the
presence of the $\exp[\omega_{2m}(\nu)\, y]$ term. Namely, for
large $m$ one has \be \omega_{2m}(\nu) \leq \omega_{2m}(\nu=0) =
{6\as \over \pi} [{-\log (|m|) - \gamma_E }]  + O(1/m^2),
\label{eq:omas} \ee which bounds from above the suppression factor
in the sum over $m$ to be $|m|^{-6 y \as / \pi}$.  This changes
the $1/|m|$ asymptotics of the summed terms and ensures the
convergence of the infinite sum. Note, that for small $y$ (i.e.
$z\to 0$) the limiting behaviour of the amplitude is \be A(y,t)
\sim  \sum_{m=1} ^\infty {8\pi \over q^2} m^{-1-4z} = {2\pi \over
z q^2} + {\rm regular \; terms}. \label{eq:18} \ee Thus we have
reproduced the answer (\ref{eq:regg2},\ref{eq:univ}), which was
obtained by accounting for the gluon reggeization when the
infrared cutoff $k_0 \to 0$.

\begin{figure}[t]
\begin{center}
\mbox{\epsfig{figure=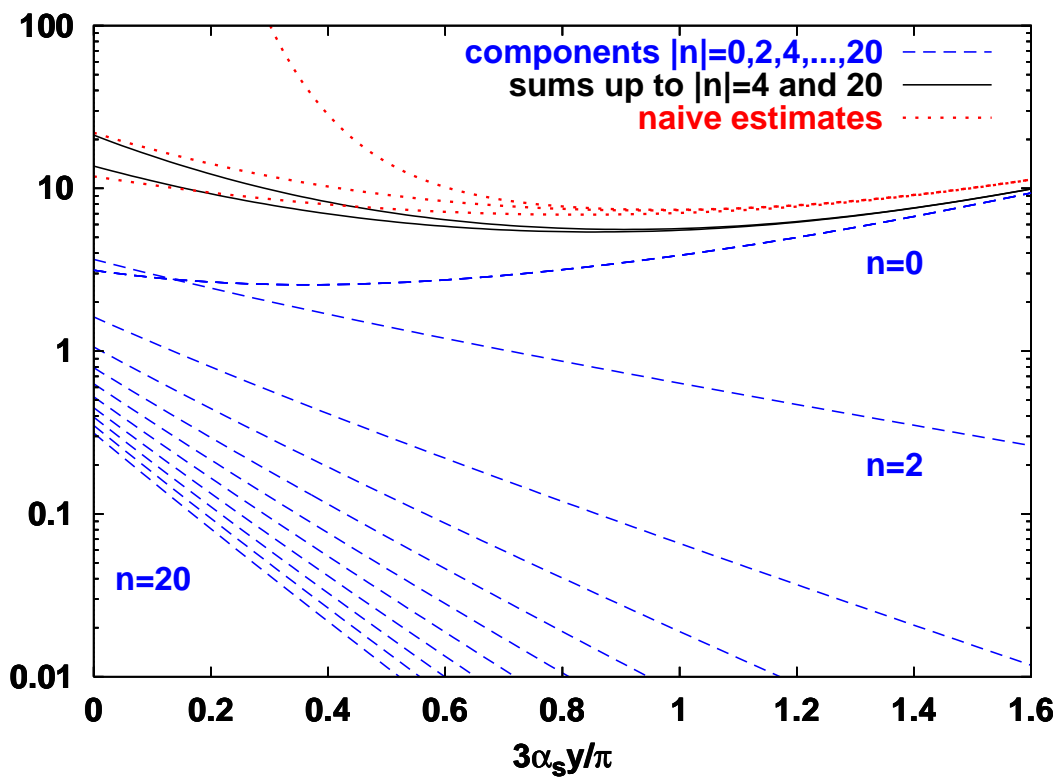,width=15cm}}
\caption{\small\em The integrals over $\nu$ in (\ref{eq:final}) corresponding to values of
$m=|n|/2$ between $0$ and $10$ (dashed lines), the sum over the integrals from
$n=-4$ to~$4$ and from $n=-20$ to~$20$ (continous lines)
and the naive sum of $n=0$ component and the reggeization term (\ref{eq:univ}) (its
l.h.s.) multiplied by $q^2/4$, plotted as functions of $3\as y / \pi$
in three cases: $q/{k_0}=4^2$, $20^2$ and ${k_0}=0$ (dotted lines).
} \label{eqn:fig1}
\end{center}
\end{figure}


In Fig.~2 we illustrate the result of taking a finite number of
terms in the sum over the conformal moments in (\ref{eq:final}).
We first plot the sum of terms with all conformal spins $n$
between -4 and 4, and then show the sum for terms between -20 and
20. Only even values of conformal spins contribute.

\subsection{Insight from comparison with naive estimates}

At very high rapidity the amplitude is dominated by the $n=0$
component, while at low rapidity it is mainly two gluon exchange
(\ref{eq:regg2}).  We may therefore try to approximate the full
amplitude by the sum of these two contributions. To compare this
approximation with amplitude summed over conformal spins, we have
to understand how the infrared parameter $k_0$ reveals itself in
the BFKL expression (\ref{eq:fbfkl}) which was written in terms of
conformal eigenfunctions. It turns out that the impact factor
$I_{n,\nu}$ goes to zero for $|n| \ga n_{\rm\small max}$, where
the value of $n_{\rm\small max}$ is regulated by the ratio
$q/k_0$. \\
%

Let us focus on the integral (\ref{eq:is}) defining the function
$I^s _{n,\nu}(\tdm k,\tdm q)$ for configurations with, say, $k\ll
q$, which dominate at low $y$. In this case the typical values of
$\rho_1$ and $\rho_2$ in integral (\ref{eq:is}) are  $|\rho_1| \gg
|\rho_2|$. The scale for $|\rho_1|$ is set by the size $R$ of the
initial colour multipole.

Then, one may rewrite 
\be 
\left( {\rho_1 - \rho_2 \over \rho_1 \rho_2} \right)^h 
\left( \left( {\rho_1 - \rho_2 \over \rho_1 \rho_2} \right)^* \right)^{\tilde h}
\simeq
\left( {1 \over \rho_2 } \right) ^{h} \left( {1 \over \rho_2^*}\right) ^{\tilde h}
\left( 
 1 - h\, {\rho_2 \over \rho_1 } - \tilde h\, {\rho_2^* \over \rho_1^* }  
\right)  
\label{eq:19b}
\ee 
where only the leading correction in $|\rho_1/\rho_2|$ to the 
result obtained in the limit $R \to \infty$, is retained. 
Thus, a rough estimate of the relative correction to (\ref{eq:ifinal}) 
coming from the fact that the size $R$ is finite, from the term proportional 
to $\;h \rho_2 / \rho_1\, $, is given by the ratio 
\be
C_{n,\nu}  =
{
{-\int d^2 \rho_2\;\, (h \rho_2 / R)\;\, \exp(iq \rho_2) \; 
\rho_2 ^{-h} (\rho_2 ^*)^{-\tilde h}} 
\over
\int d^2 \rho_2\; \exp(iq \rho_2) \; \rho_2 ^{-h} (\rho_2 ^*)^{-\tilde h}}.
\label{eq:CF}   
\ee
The integrals in (\ref{eq:CF}) are of the same form as those used for the derivation
of (\ref{eq:ifinal}). In particular, the result for the denominator 
is already known to be
\be
{2\pi\, i^n \over q} \;  \left( {q \over 2 } \right)^{2i\nu}
{\Gamma(1/2+ n/2 -i\nu) \over \Gamma(1/2 + n/2 + i\nu)}
\label{eq:denom}
\ee
and the integral in the numerator may be obtained from 
(\ref{eq:denom}) by substitutions $i\nu \to i\nu - 1/2$
and $n \to n-1$. Thus, it is straightforward to find that
\be
C_{n,\nu} = {(n/2+i\nu)^2 - 1/4 \over iqR}. 
\label{eq:CF2}
\ee
It may be seen that the typical value of $|\nu|$  
in integral (\ref{eq:fbfkl}) is $|\nu| \simeq |n|/2$ 
for large $|n|$. 
Then $|(n/2+i\nu)^2 - 1/4| \simeq n^2 /2$  and after including
a similar result from the term in eq.~(\ref{eq:19b}) 
containing $\;\tilde h {\rho_2^*  /\rho^* _1}\;$, 
the estimate of the total correction is  
\be
2|C_{n,\nu}| \simeq {n^2\over qR}. 
\label{eq:2c}
\ee
The Mueller-Tang scheme breaks down when the correction factor becomes 
large $2|C_{n,\nu}| \ga 1$ and $I^s_{n,\nu}$ gets
suppressed in relation to the r.h.s of (\ref{eq:ifinal}). 
Noting that ${k_0 ^2} \simeq 1/R^2$, we find from (\ref{eq:2c}) 
that the sum over conformal spins should be extended to 
\be 
n_{\rm\small max} = \sqrt{{ q \over k_0}}. 
\label{eq:nmax} 
\ee

We may compare the result of the summation up to $n_{\rm\small max}$ 
with naive estimates given by a simple addition of the $n=0$
contribution and the amplitude given by the two reggeized gluon
exchange (\ref{eq:regg2}). The dashed curves in Fig.~2 show
results for three values of the ratio $q/{k_0}$, namely $4^2$,
$20^2$ and $\infty$. The agreement between the naive estimates and
the truncated sum over conformal spins can be understood as
follows.
At $y=0$ the sum over $n$ in (\ref{eq:fbfkl}) takes form \be
A(y=0,t) \simeq {8\pi \over q^2} \left( {1\over 2}+\sum_{m=1}
^{n_{\rm\tiny max}/2} {1\over m}  \right) \simeq {2\pi \over q^2}
\log \left( {q^2 \over k_0^2} \right), \ee where we have used
(\ref{eq:final}) and (\ref{eq:y0}). In this way we reproduce the
leading behaviour of the amplitude given by (\ref{eq:log}).
Indeed, as seen in Fig.~2, the naive sum of the $n=0$ component
and the two (reggeized) gluon contribution (\ref{eq:regg2}) for
the two cases of $q/k_0=4^2$ and $20^2$ are rather close to the
sum over the conformal spins $|n|$ up to 4 and 20
respectively\footnote{Note that, in general, the naive sum
slightly overestimates the true result as the two gluon
contribution (\ref{eq:regg2})
already contains part of the $n=0$ component.}.\\

Note that we do not have any non-perturbative effects modifying
the gluon perturbative propagators.
However, the sensitivity to the infrared details decreases with
increasing $y$, since these details influence more the
contributions from higher conformal spins, which get suppressed
with $y$.  The suppression is faster for larger $|n|$.
In view of the above discussion and the recent results of
Ref.~\cite{EMI,EIM}, the extension to include higher conformal spins
is necessary to understand the Tevatron data for hard colour
singlet exchange.

\section{Phenomenological consequences for gaps between jets}

As we have just discussed, the above formalism is relevant to
processes mediated by colour singlet exchange with high momentum
transfer $q$. As noted, the classic example is the production of a
pair of high $E_T$ jets separated by large rapidity gap $y$. We
emphasize that the conventional (asymptotic) BFKL amplitude only
dominates at high rapidities well above the reach of present
experiments. This is illustrated by the discrepancy between the
$n=0$ curve and the full result (given by the continous curves on
Fig.~2), in the region $3\as y /\pi < 1$ currently sampled
experimentally. Instead, in this region it is necessary to sum all
the contributions with conformal spins up to $n_{\small\rm
max}=\sqrt{q/k_0}$, where $k_0$ is a physical infrared cutoff. On
the other hand, in this domain the amplitude is well described by
two reggeized gluon exchange, as illustrated by the dotted
curves in Fig.~2 (note, that the uppermost curve corresponds
$q/k_0 \to \infty$ whereas the lower dotted curves correspond
to the choices $\sqrt{q/k_0}=4$ and~20). The physical meaning of
gluon reggeization, that is of the factor $(k^2/q^2)^z$ in
(\ref{eq:regg},\ref{eq:regg2}), is that it reflects the fact that
the emissions of extra gluons are forbidden within the rapidity
gap interval $y$, so that  pure elastic parton-parton scattering
occurs. This is equivalent to the normal Sudakov-like suppression.
The latter suppression is the probability not to emit gluons with
transverse momentum $p_t$ in the interval $(k_0,q)$, which takes
the form $\exp (-n_g)$.  The quantity $n_g$, the anticipated
average number of emissions, is \be n_g = \int_{k_0^2} ^{q^2} {d
p^2_t \over p_t ^2} {3 \as \over \pi} y  . \label{eq:ng} \ee Thus,
the non-forward lowest-order two-gluon exchange {\it amplitude}
should be multiplied by the suppression factor \be \exp(-n_g/2) =
\left( {k_0^2 \over q^2} \right)^{3\as y / (2\pi)}, \ee as given
in (\ref{eq:regg}). Thus we see that in the BFKL amplitude the
suppression is generated by the resummation of the virtual
corrections.

\begin{figure}[t]
\begin{center}
\mbox{\epsfig{figure=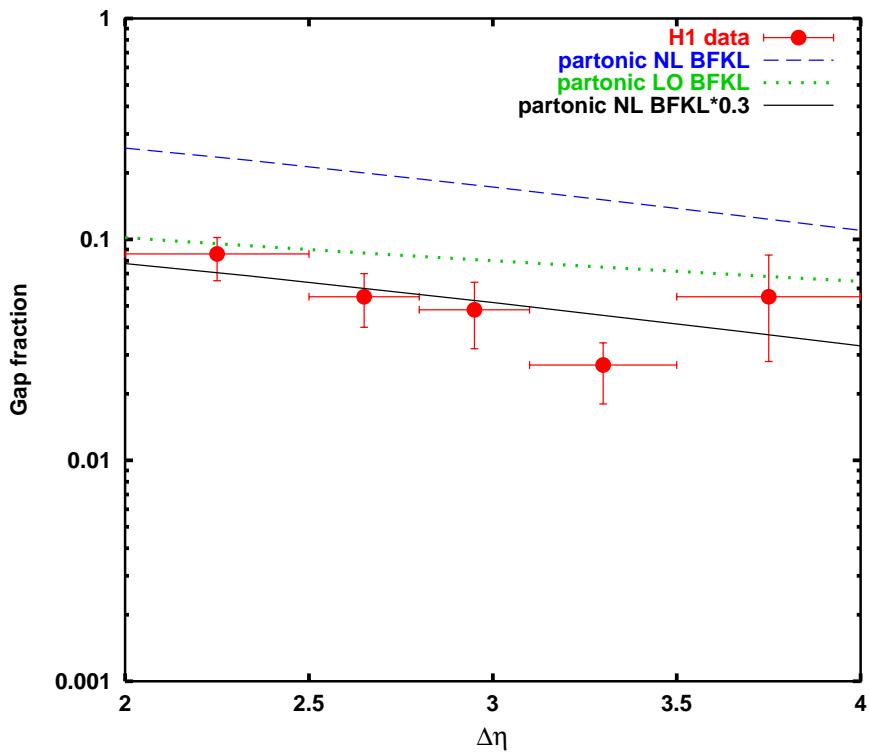,width=12cm}}
\end{center}
\caption{\small\em
Gap fraction for photoproduction of two high $E_T$ jets separated by a rapidity gap
$\Delta\eta$. The experimental data \cite{H1} are obtained with the rapidity gap events
defined as those with the maximal total transverse energy flow in between the 
high-$E_T$ jets to be $E_T ^{\rm\small cut} = 0.5$~GeV.
The continous and dashed curves show the partonic non-leading BFKL prediction, 
with and without the gap survival probability factor included respectively,
and the dotted curve is the LO BFKL prediction without the 
gap survival probability factor taken into account.
 }
\end{figure}


The hard colour singlet exchange has been investigated experimentally at the 
Tevatron \cite{TEVEXP} by measuring events with gaps between jets. 
In \cite{CFL} these data were compared with
BFKL results in the standard Mueller-Tang approximation. It was found
that the rising $E_T$ dependence of the gap fractions may be reproduced
by the model only when a fixed value of coupling constant is used.
The assumed lack of running of the coupling with increasing 
momentum transfer is, however, difficult to motivate. 
The experimental status of the gap fraction dependence on the jet separation in 
rapidity, $\Delta y$, is not so certain as the observed $E_T$ dependence. 
In particular, CDF results indicate a decreasing tendency at high  $\Delta y$, 
contrary to D0 data. The error bars are still too large to claim inconsistency
and both the experimental distributions agree with a flat $\Delta y$ dependence.
This is, however, incompatible with the predictions based on the
Mueller-Tang approximation, which give a steep rise of the gap fraction
at large rapidity \cite{CFL}.

In the Tevatron kinematical conditions the bulk of data comes from 
$E_T \sim 20$~GeV and $\Delta y \sim 5$. 
Then, if our choice of the cutoff scale $k_0 \sim 1$~GeV$^2$ is correct, 
the Mueller-Tang approximation gives about a half of the contribution 
to the scattering amplitude, thus about 25\% of the cross-section.
After setting $\alpha_s =0.17$, as in \cite{CFL},
this may be seen in Fig.~2 by comparing the sum over conformal spins
up to $n=4$ with $n=0$ component, at $3\alpha_s y /\pi \sim 0.8$.
Therefore the Mueller-Tang approximation should not be used to 
describe the available Tevatron data unless the cut-off $k_0$
is much larger than we expect. 

Recently \cite{EMI,EIM} it has been demonstrated that
when the gluon reggeization phenomenon is accounted for, no
significant discrepancies between the data and BFKL results appear 
in either the $E_T$ or the $\Delta y$ distribution.
This conclusion holds both for fixed and running couplings, provided they are  
consistently used in the amplitude at the scale set by the
typical virtuality of each vertex.

The result obtained in Refs.~\cite{EMI,EIM}, from theoretical considerations based 
on the BFKL framework, was confronted with experimental data from the CDF and D0 
collaborations \cite{TEVEXP}.
Besides the hard parton-parton scattering amplitude for colour singlet exchange, 
other important effects, like hadronisation corrections and the gap survival probability
were included using a complete Monte Carlo treatment. 
It was found \cite{EMI,EIM} that the full BFKL prediction was in 
agreement with the Tevatron data, contrary to calculations based on the asymptotic
Mueller-Tang approximation.


An interesting way of studying this effect in more detail is to
observe events with a large rapidity gap between two high $E_T$
jets in diffractive deep inelastic scattering at HERA, as sketched
in Fig.~1. Since one jet comes from photon dissociation, the
infrared cutoff $k_0$ is controlled by the photon virtuality
$Q^2$. Therefore, there is the possibility to vary $k_0$ while
retaining the {\em same} kinematics of the hard parton-parton
interaction. In this way, we avoid complications from
hadronisation, variation of parton densities etc. Another
advantage of {\it electroproduction} is that we have high survival
probability, $S^2$, of the rapidity gap against soft
rescatterings, that is $S^2 \approx 1$, contrary to the analogous
jet production process at the Tevatron.

At present, however, data are available at HERA \cite{H1} for the {\it photoproduction} of 
jets separated by a rapidity gap. 
These data are compared with the corresponding non-forward BFKL
predictions in Fig.~3. The two uppermost curves 
are obtained using the solutions to the BFKL equation given in
Ref.~\cite{EMI,EIM} and correspond to BFKL amplitudes with and without
incorporating resummations of higher order effects. 
In the leading order BFKL calculation a fixed value of $\alpha_s=0.17$
was used, whereas the running coupling was taken in the non-leading BFKL 
case, which explains why the LO BFKL curve lies so low.  
Nevertheless,
for {\em photoproduction} the probability of soft rescattering,
which produces secondaries in the rapidity gap, is not negligible.
To calculate the resulting suppression factor $S^2$ we have used
the formalism of Ref.~\cite{KMR}. Recall that, there, the model
was tuned to describe the available soft $pp$ and $p\bar p$
interactions throughout the CERN ISR--Tevatron energy range. Assuming
Vector Meson Dominance, the rescattering in photoproduction may
occur between a virtual vector meson and the proton. Data are not
available for soft vector meson--proton scattering. However using
the additive quark model, and the analogy between the light vector
mesons $V=\rho,\omega,...$ and the pion, we expect\footnote{Note
that the ZEUS collaboration \cite{ZEUSpi} estimate that
$\sigma_{\rm\small tot} (\pi p) = 31 \pm 4$~mb, by observing the
interference between the pions from $\rho$ meson and $\pi^+ \pi^-$
background.} \be \sigma_{\rm\small tot} (Vp) \simeq 30~\mbox{\rm
mb} \label{eq:svp} \ee at the collision energies relevant to the
HERA data, that is $W \simeq 200$~GeV. With this cross section we
determine the relevant photoproduction rapidity gap survival
probability to be \cite{KMR}\footnote{Inputting the cross section
of (\ref{eq:svp}), the model \cite{KMR} predicts the elastic slope
$B \simeq 11$~GeV$^{-2}$, in agreement with ZEUS observations for
$\gamma p \rightarrow \rho p$.} \be S^2 \simeq 0.3. \ee After
taking this factor into account we obtain the final prediction
shown by the lower continuous curve in Fig.~3. There is thus an
excellent agreement between the data and the theory.

It is appropriate to comment on the approximations made to obtain this prediction 
for the {\it photoproduction} of jets separated by a rapidity gap at HERA.  
We have assumed that the effects of hadronization and of producing gaps in the 
conventional colour-octet exchange scattering are small.  
We expect the experimental cut $E_T^{\rm cut}$ on soft secondaries to suppress 
these  effects.  

Moreover, although we believe our two-channel eikonal calculation of the 
survival factor $S^2$, using the framework of \cite{KMR}, is the best that can be done at 
present, we note that again it relies on soft phenomena.  Clearly it is important to study 
all these effects in more detail within a Monte Carlo framework as in 
\cite{EMI,EIM}. This will, among other things, allow a study of the influence 
of the parameter $E_T^{\rm\small cut}$ in the gap definition \cite{Sterman}
on the gap fraction.
  
When the HERA luminosity increases, it will be particularly informative to
observe {\em electroproduction} of high $E_T$ jets separated by
large rapidity gaps. This will allow $Q^2$, as well as the jet $E_T$, to
be varied, and hence conformal spin summations up to different
$n_{\rm\small max}$ to be probed.  Moreover, here, there are no uncertainties 
connected with the survival factor $S^2$, since it is predicted to be 1 for this process.  
These data will therefore allow a test of QCD radiative effects, which are 
important ingredients in all non-forward diffractive phenomena.

\section*{Acknowledgements}

We thank G.~Ingelman and R.~Enberg for useful comments. LM is grateful to the Grey College
and the Department of Physics of the University of Durham for their hospitality and the 
Swedish Natural Science Research Council for the fellowship.
This work was supported in part by the UK Particle Physics and Astronomy Research Council, 
the Russian Fund for Fundamental Research (grants 01-02-17095 and 00-15-96610), 
the EU Framework TMR programme, contract FMRX-CT98-0194 (DG 12-MIHT) and 
by the Polish State Committee for Scientific Research grant 5~P03B~144~20.

\end{document}